        \documentstyle[11pt]{article}
        \newcommand{\del}{\delta}
        \newcommand{\eps}{\epsilon}
        \newcommand{\kap}{\kappa}
        \newcommand{\lam}{\lambda}
        \newcommand{\om}{\omega}
        \newcommand{\th}{\theta}
	\newcommand{\vph}{\varphi}
        \newcommand{\Om}{{\it \Omega}}
        \newcommand{\Gam}{{\it \Gamma}}

        \newcommand{\EQNO}[1]{\setcounter{equation}{0}
                              \def\theequation{{#1}.\arabic{equation}}}
        \newcommand{\SECT}[2]{\EQNO{{#1}}
                \vspace{4.5ex}
                \begin{center}
                {{#1}. {\bf {#2}}}
                \end{center}}
	\newcommand{\sect}[1]{Sect.$\,${#1}}
        \newcommand{\be}{\begin{equation}}
        \newcommand{\ee}{\end{equation}}
        \newcommand{\bea}{\begin{eqnarray}}
        \newcommand{\eea}{\end{eqnarray}}
        \newcommand{\nno}{\nonumber \\}
        \newcommand{\eqlb}[1]{\label{eq: {#1}}}
        \newcommand{\eqrf}[1]{($\!\!$~\ref{eq: {#1}})}
        \newcommand{\sep}[1]{\!\!\!\!\! &{#1}& \!\!\!\!\! }
        \newcommand{\eq}{\sep{=}}
        \newcommand{\vc}{\sep{ }}
        
        \newcommand{\lo}{\mbox{\LARGE ( } \!\!\!}
        \newcommand{\lc}{\mbox{\LARGE ) } \!\!}
	\newcommand{\ind}[1]{\!\!^{^{^{{\scriptstyle {#1}}}}}}
	\newcommand{\expl}[1]{{\mbox{[{#1}]}}\hspace{-10em}}
	\newcommand{\thm}[2]{\vspace{3ex} \noindent {\bf {#1}.} {#2}}
	\newcommand{\pf}[1]{\noindent {\em Proof.} {#1} $\;\;\Box$\vspace{2ex}}

        \newcommand{\vol}{{\rm vol}\/}
        \newcommand{\sgn}{{\rm sgn}}
        \newcommand{\e}{e}
        \newcommand{\dr}{d}
        \newcommand{\pdr}{{\partial}}
        \newcommand{\inv}[1]{\frac{1}{#1}}
        \newcommand{\hf}{\inv{2}}
	\newcommand{\im}{{\rm Im}}

        \font\bbb=msym7
        \font\BB=msym10
        \font\BBB=msym10 at 11 pt
        \font\sfrak=eufm7
        \font\frak=eufm10 at 11 pt

	\newcommand{\mm}{moment map }
        \newcommand{\dh}{Duistermaat-Heckman }
        \newcommand{\tdr}{\tilde{\dr}}
        \newcommand{\re}{{\mbox{\bbb{R}}}}
        \newcommand{\RE}{{\mbox{\BBB{R}}}}
        \newcommand{\CO}{{\mbox{\BBB{C}}}}
        \newcommand{\co}{{\mbox{\bbb{C}}}}
	\newcommand{\NA}{{\mbox{\BBB{N}}}}
        \newcommand{\ZE}{{\mbox{\BBB{Z}}}}
        \newcommand{\ze}{{\mbox{\BB{Z}}}}
        \newcommand{\gt}[1]{{\mbox{\frak{{#1}}}}}
        \newcommand{\g}{\gt{g}}
        \newcommand{\sg}{{\mbox{\sfrak{g}}}}
	\newcommand{\tpe}{\inv{(2\pi\eps)^\hf}}
	\newcommand{\hfn}{{[\frac{n-1}{2}]}}
	\newcommand{\frn}{\frac{(2\pi)^{n-1}}{(n-1)!}}
	\newcommand{\de}[1]{\frac{\dr {#1}}{(2\pi\eps)^\hf}\,}
	\newcommand{\df}{\frac{\dr\phi}{2\pi}\,}
	\newcommand{\mi}[1]{\mu^{-1}({#1})}
	\newcommand{\pdo}[1]{\frac{\pdr}{\pdr {#1}}}
	\newcommand{\ep}[1]{\e^{-\frac{{{#1}}^2}{2\eps}}}
	\newcommand{\err}{o(\ep{\del})}
	\newcommand{\pap}{\prod_{i=1}^na_i(p)}
	\newcommand{\pan}{\prod_{i=1}^{n_N}a_i(N)}
	\newcommand{\app}{\inv{a_1(p)a_2(p)}}
	\newcommand{\ann}{\inv{a_1(N)}}
	\newcommand{\ipn}{(i\phi)^n}
	\newcommand{\ipm}{i\phi\mu}
	\newcommand{\hfe}{\frac{\eps}{2}}
	\newcommand{\invp}{\inv{2\pi}}
	\newcommand{\invg}{\inv{|\Gam_0|}}
	\newcommand{\sump}{\sum_{p\in F}}
	\newcommand{\sumn}{\sum_{N\subset F}}
	\newcommand{\sumk}{\sum_{k=0}^\hfn}
	\newcommand{\sumkn}{\sum_{k=n_N}^n}
	\newcommand{\sumi}{\sum_{i=1}^n}
	\newcommand{\prdi}{\prod_{i=1}^n}
	\newcommand{\lv}{\frac{\om^n}{n!}\,}
	\newcommand{\lvl}[2]{\frac{\om_{{#1}}^{{#2}}}{({#2})!}}
	\newcommand{\smp}{\sgn\mu(p)}
	\newcommand{\di}{(-\del, \del)}
	\newcommand{\inti}{\int_0^\infty}
	\newcommand{\intii}{\int_{-\infty}^\infty}
	\newcommand{\euler}{e_{S^1}(\nu_N)}
	\newcommand{\comb}[2]{{{\scriptstyle \left(
				\begin{array}{c}
				{#1} \\ {#2}
				\end{array}
				\right)}}\,}
	\newcommand{\two}[4]{\left\{	\begin{array}{ll}
					{#1}, & {\mbox{if }} {#2}, \\
					{#3}, & {\mbox{if }} {#4}
					\end{array}	\right.}
	\newcommand{\three}[6]{	\left\{	\begin{array}{ll}
				{\textstyle {#1}}, & {\mbox{if }} {#2}, \\
				{\textstyle {#3}}, & {\mbox{if }} {#4}, \\
				{\textstyle {#5}}, & {\mbox{if }} {#6}
					\end{array}	\right.}

\begin{document}
$\!\,{}$

	\vspace{-5ex}

	\begin{flushright}
{\tt hep-th/9212071}\\
December, 1992
	\end{flushright}

        \begin{center}
        {\LARGE\bf An Integration Formula for the \\
	\vspace{1ex}
			Square of Moment Maps of Circle Actions}\\
        \vspace{3ex}

        {Siye Wu\footnote{e-mail address: {\tt sw@shire.math.columbia.edu}}}\\
	\vspace{1ex}
	{\small Department of Mathematics\\
	Columbia University\\
	New York, NY 10027}
        \end{center}

	\vspace{2ex}

\begin{quote}
{\small
The integration of the exponential of the square of the moment map of
the circle action is studied by a direct stationary phase computation
and by applying the Duistermaat-Heckman formula.
Both methods yield two distinct formulas expressing the integral in terms
of contributions from the critical set of the square of the moment map.
The cohomological pairings on the symplectic quotient, including its volume
(which was known to be a piecewise polynomial), are computed explicitly
using the asymptotic behavior of the two formulas.}
\end{quote}

        \SECT{1}{Introduction}

The formula of \dh [\ref{DH}] is an exact stationary phase formula
for integrating the exponential of the moment map.
It was put into and proved in the context of equivariant cohomology
in [\ref{BV}] and [\ref{AB}].
The formula has been applied in [\ref{GP}] to the case
of non-abelian group actions under certain regularity assumptions.
More recently, with the idea of ``non-abelian localization'',
Witten [\ref{W}] discovered a remarkable integration formula
for the exponential of the square of the moment map.
An infinite dimensional version of this formula agrees completely with
the partition function of two dimensional quantum Yang-Mills theory
that can be computed independently by more physical methods [\ref{W}].
{}From the point of view of symplectic geometry, a precise expression
of the integral would certainly be interesting
and may well have significant applications.
The purpose of this paper is to study a special case of circle actions
in detail.

First of all, let me recall the argument of Witten [\ref{W}]
on why such a formula should exist.
Given a Hamiltonian action of a compact connected Lie group $G$ (with Lie
algebra $\g$) on a compact connected symplectic manifold
$(M, \om)$ of dimension $2n$.
Let $\mu\colon M\to\g^*$ be the \mm, satisfying $i_{X_a}\om=\dr\mu_a$
for $a=1, \cdots, m$, $(m=\dim G)$, where $X_a$ are the vector fields on $M$
induced by the elements of a fixed chosen basis of $\g$.
The invariant inner product on $\g$ induces one on its dual $\g^*$,
both of which are denoted by $(\cdot, \cdot)$. Consider the integral
        \be
Z(\eps)=\inv{\vol(G)}\inv{(2\pi\eps)^{\frac{m}{2}}}
\int_M\lv\e^{-\inv{2\eps}I},			\eqlb{Z}
        \ee
where $I=(\mu, \mu)$ is the square of the moment map.
Introducing a Gaussian integration, this integral is equal to\footnote{We
sum over repeated indices of the Lie algebra unless stated otherwise.}
        \be
Z(\eps)=\inv{\vol(G)}\int_\sg\frac{\dr^m\phi}{(2\pi)^m}
\,\e^{-\hf\eps(\phi,\phi)}\int_M\e^{\om+i\phi^a\mu_a}.
        \ee
The integral over $M$ reminds us of the standard \dh formula, except
the group $G$ could be non-abelian now.
Since $\tilde{\om}=\om+i\phi^a\mu_a$ is the equivariant extension of $\om$,
$\e^{\tilde{\om}}$ is (formally) an equivariantly closed form.
Here the differential acting on the space of equivariant differential forms
$\Om_G^*(M)=(\Om^*(M)\otimes S(\g^*))^G$ is $\tdr=\dr-i\phi^ai_{X_a}$
[\ref{MQ}]. For any real number $t$ and any $\lam\in\Om_G^*(M)$, we have
        \be
\int_M\e^{\tilde{\om}}=\int_M\,\e^{\tilde{\om}-t\tdr\lam},	\eqlb{dheq}
        \ee
because the difference of the two integrands is equivariantly exact. Therefore
        \be
Z(\eps)=\inv{\vol(G)}\int_\sg\frac{\dr^m\phi}{(2\pi)^m}
\,\e^{-\hf\eps(\phi,\phi)}\int_M\,\e^{\tilde{\om}-t\tdr\lam}.	\eqlb{Zeq}
        \ee
The right hand side is independent of $t$. On the other hand, apart from
a polynomial in $t$ that comes from the expansion of $\e^{-t\dr\lam}$,
$t$ appears entirely in a factor $\e^{it\phi^ai_{X_a}\lam}$.
The methods of stationary phase implies that as $t\to\infty$, the integral
\eqrf{Zeq} is concentrated on the critical set of $\phi^ai_{X_a}\lam$,
a subset of $\g\times M$ satisfying $i_{X_a}\lam=0$ and
$\phi^a\dr i_{X_a}\lam=0$.
The contribution to the integral \eqrf{Z} from any compact subset of
$\g\times M$ disjoint with the critical set tends to zero as $t\to\infty$.
In the symplectic case, pick a positive $G$-invariant almost complex structure
$J$ on $(M, \om)$ and let $\lam=\hf J(\dr I)$.
Then $i_{X_a}\lam=0$ if and only if $\lam=0$ [\ref{W}].
Therefore apart from the $\phi$-part, the critical set of $\phi^ai_{X_a}\lam$
coincides with that of $I=(\mu, \mu)$.
If each connected components $N$ of the critical set contributes
$Z_N(\eps)$ to \eqrf{Z}, then we have [\ref{W}]
        \be
Z(\eps)=\sum_NZ_N(\eps).	\eqlb{ZN}
        \ee
The precise expression for $Z_N(\eps)$, though not yet known in the most
general cases, has been computed explicitly if a neighborhood of $N$
can be modeled on the cotangent bundle of the group $G$ or its
homogeneous spaces [\ref{W}].

If $G=S^1$ acts on a symplectic manifold $(M, \om)$
in a Hamiltonian fashion, then the \mm $\mu\colon M\to\RE$
is a Morse function (in the sense of Bott) with many special properties.
For example, the critical set $F$ of $\mu$ is a symplectic submanifold
of $M$ and for any regular value $\xi\in\RE$ of $\mu$, the level set
$\mi{\xi}$ is a connected submanifold of $M$ [\ref{A}].
These features facilitate the computation [\ref{W}] described above,
yielding a precise expression of $Z_N(\eps)$.
On the other hand, since $G=S^1$, the standard technique of abelian
localization [\ref{AB}] is available and can be applied readily to a
more rigorous proof of the formula.
What was not expected before is that $Z_N(\eps)$ in \eqrf{ZN}
is not unambiguously defined.
In the first approach, it depends on whether $t\to+\infty$ or
$t\to-\infty$ (although the total contribution $\sum_N$ doesn't),
whereas in the second one, the $\phi$-integration for a single
component $N$ depends on the choice of contour in the complex $\phi$-plane.
Consequently, there are two formulas realizing \eqrf{ZN}.

In \sect{2}, we find explicit forms of \eqrf{ZN} for $G=S^1$
using Witten's method [\ref{W}].
The rest of the paper is devoted to a more traditional proof and
the applications of this result.
Most of the discussions generalize straightforwardly to the case of
torus actions.
We leave this and the non-abelian case for future investigation.

	\SECT{2}{Stationary Phase Computation}

Let $\mu\colon M\to\RE$ be a \mm of the Hamiltonian $S^1$ action.
The stationary phase method reduces the integral \eqrf{Zeq} to one on
the critical set of $\mu^2$. This set is the union of the critical set
and the zero set of $\mu$.
The former is also the fixed-point set $F$ of the $S^1$ action.
We assume that $F$ is discrete.
Near a point $p\in F$, we can use the linear coordinates
$\{(q_i,p_i),i=1,\cdots,n\}$ on $T_pM$ compatible to $\om$ and $J$, i.e.,
$\om=\sumi\dr q_i\wedge\dr p_i$ and $J\dr q_i=-\dr p_i$, $J\dr p_i=\dr q_i$.
Let $a_i=a_i(p)$ ($i=1,\cdots,n$) be the (non-zero) weights\footnote{We
use the convention $a_i\in\ze$.} of the isotropic representation
of $S^1$ on $T_pM$, then the vector field induced by the $S^1$ action
is $X=\sumi a_i(q_i\pdo{p_i}-p_i\pdo{q_i})$ and the \mm
is $\mu=\mu_0-\hf\sumi a_i(q_i^2+p_i^2)$, up to higher order terms.
Assuming $\mu_0\neq 0$, a direct computation shows that
$\lam=\hf J\dr\mu^2=\mu_0\sumi a_i(q_i\dr p_i-p_i\dr q_i)+\cdots$,
and the equivariant form appeared in \eqrf{Zeq} is,
again up to higher order terms,
        \bea
\tilde{\om}-t\tilde{\dr}\lam
\eq\om+i\phi\mu-t\,(\dr\lam-i\phi\,i_X\lam)		\nno
\eq\sumi(1-2\mu_0a_it)\,\dr q_i\wedge\dr p_i+i\phi\mu_0
   -\hf\,i\phi\sumi a_i(1-2\mu_0a_it)(q_i^2+p_i^2).
        \eea
As $t\to\pm\infty$, the contribution from $p\in F$ can be
computed by Gaussian approximation [\ref{W}]:
        \bea
Z_p(\eps)
\eq\invp\int_\re\df\e^{-\hfe\phi^2}
   \int_{\re^{2n}}\prdi(1-2\mu_0a_it)\dr q_i\wedge\dr p_i
   \,\e^{i\phi[\mu_0-\hf\sumi a_i(1-2\mu_0a_it)(q_i^2+p_i^2)]}    \nno
\eq\invp\tpe\int_{\re^{2n}}\prdi(1-2\mu_0a_it)\dr q_i\wedge\dr p_i
   \,\e^{-[\mu_0-\hf\sumi a_i(1-2\mu_0a_it)(q_i^2+p_i^2)]^2/2\eps}
        \eea
If $t\to+\infty$, let\footnote{We define the function $\sgn(x)=x/|x|$
for $x\neq 0$.} $s_i=\hf(-\sgn\mu_0)a_i(1-2\mu_0a_it)(q_i^2+p_i^2)$, then
        \bea
Z_p^+(\eps)
\eq\invp\tpe\inti(2\pi)^n\frac{(-\sgn\mu_0)^n}{\prdi a_i}
   \prdi\dr s_i\,\ep{(\sumi s_i+|\mu_0|)}                         \nno
\eq\frn\frac{(-\sgn\mu_0)^n}{\prdi a_i}\inti\de{s}s^{n-1}\ep{(s+|\mu_0|)}.
	\eqlb{Zp+}
	\eea
If however $t\to-\infty$, let
$s_i=\hf(\sgn\mu_0)a_i(1-2\mu_0a_it)(q_i^2+p_i^2)$, then
        \be
Z_p^-(\eps)=\frac{(\sgn\mu_0)^n}{\prdi a_i}
\frn\inti\de{s}s^{n-1}\ep{(s-|\mu_0|)}.			\eqlb{Zp-}
        \ee
Indeed the two cases $t\to+\infty$ and $-\infty$ do not yield same results.
Particularly important is their different asymptotic behavior as $\eps\to+0$.
Let $\del=\min\{|\mu(p)|, p\in F\}$.
Since $|\mu_0|\ge\del$, a straightforward calculation shows that\footnote{Here
$\err$ stands for any $f(\eps)$ such that as $\eps\to+0$,
$|f(\eps)|\,\e^{\del^2/2\eps}\le C$, a fixed constant.}
	\be
\inti\de{s}s^{n-1}\ep{(s+|\mu_0|)}=\err
	\ee
whereas
	\bea
\vc\inti\de{s}s^{n-1}\ep{(s-|\mu_0|)}					\nno
\eq\intii\de{s}s^{n-1}\ep{(s-|\mu_0|)}+\err				\nno
\eq\sumk\comb{\! n-1\!}{2k}|\mu_0|^{n-1-2k}\intii\de{s}s^{2k}\ep{s}+\err\nno
\eq\sumk\frac{(n-1)!}{(n-1-2k)!2^kk!}\eps^k|\mu_0|^{n-1-2k}+\err.
	\eea

Next we study the zero set $\mi{0}$.
Let $\th$ be the coordinate in the $S^1$ direction
and let $\xi$ be the value of the \mm $\mu$.
Then $\th$, $\xi$ and a (local) coordinate on the quotient $M_0=\mi{0}/S^1$
provides a coordinate on a neighborhood of $\mi{0}$.
Since $\xi$ is the conjugate variable of $\th$, the symplectic form
$\om$ near $\mi{0}$ must contain the term $\dr\th\wedge\dr\xi$.
One should be more cautious about the definition of $\th$.
If $S^1$ acts freely on $\mi{0}$, then $\mi{0}\to M_0$ is a principal
$S^1$-bundle and is in general non-trivial (see \sect{6}).
A local section $s$ over an open set $U\subset M_0$ provides a trivialization
of the bundle over $U$ and hence the coordinate $\th$.
For a different section $s'$ related to $s$ by $s'=s+\vph$,
where $\vph$ is an $S^1$-valued function on $U$,
the corresponding coordinate $\th'$ is related to $\th$ by
a gauge transformation $\th'=\th-\vph$.
As a consequence, the term $\dr\th\wedge\dr\xi$ is not gauge invariant.
Introducing an $S^1$-connection on the bundle, we have a gauge potential
on $M_0$ that transforms as $A\mapsto A'=A+\dr\vph$.
Gauge invariance modifies the term $\dr\th\wedge\dr\xi$ to
$(\dr\th+A)\wedge\dr\xi$.
Since $\om$ is also a closed form and projects down to $\om_0$ on $M_0$,
we have, up to an exact 2-form on $M_0$,
	\be
\om=(\dr\th+A)\wedge\dr\xi+\xi F_0+\om_0,	\eqlb{2om}
	\ee
where $F_0=\dr A$ is the curvature.
\eqrf{2om} will be confirmed by Lemma 4.1 using a more rigorous argument.

In a neighborhood of $\mi{0}$, $X=\pdo{\th}$, $J\dr\xi=\dr\th+A$ and
$\lam=\xi(\dr\th+A)$. Hence
	\be
\tilde{\om}-t\tilde{\dr}\lam=
(1+t)(\dr\th+A)\wedge\dr\xi+(1+t)i\phi\xi+\om_0+\xi F_0.
	\ee
Using the stationary phase approximation, the contribution of $\mi{0}$
to \eqrf{dheq} is
	\bea
\vc\int_\re(1+t)\dr\xi\,\e^{i(1+t)\phi\xi}
   \int_{M_0}\e^{\om_0+\xi F_0}\int_{S^1}\dr\th+A	\nno
\eq 2\pi\sum_{k=0}^{n-1}\int_\re(1+t)\dr\xi\,\xi^k\,\e^{i(1+t)\phi\xi}
   \int_{M_0}\lvl{0}{n-1-k}\wedge\frac{F_0^k}{k!}	\nno
\eq(2\pi)^2\sum_{k=0}^{n-1}\frac{1+t}{|1+t|}\inv{i^k}\frac{\dr^k}{\dr\phi^k}
   \del(\phi)\int_{M_0}\lvl{0}{n-1-k}\wedge\frac{F_0^k}{k!},
	\eea
where $\del(\phi)$ is Dirac's $\del$-function centered at $\phi=0$.
Again, the limits of $t\to+\infty$ and $-\infty$ are different. Consequently
	\bea
   Z_{\mi{0}}^\pm
\eq\invp\int_\re\df\e^{-\hfe\phi^2}
   \sum_{k=0}^{n-1}\pm(2\pi)^2\inv{i^k}\frac{\dr^k}{\dr\phi^k}\del(\phi)
   \int_{M_0}\lvl{0}{n-1-k}\wedge\frac{F_0^k}{k!}		\nno
\eq\pm\sumk\inv{i^{2k}}\lo\frac{\eps}{2}\lc\ind{k}H_{2k}(0)
   \int_{M_0}\lvl{0}{n-1-2k}\wedge\frac{F_0^{2k}}{(2k)!}	\nno
\eq\pm\sumk\eps^k\int_{M_0}\lvl{0}{n-1-2k}\wedge\frac{F_0^{2k}}{2^kk!}.
	\eqlb{Z0}
	\eea
Here we have used the values $H_{2k}(0)=(-1)^k2^k(2k-1)!!$, $H_{2k+1}(0)=0$
of the Hermite polynomial.

Combining \eqrf{Zp+}, \eqrf{Zp-} and \eqrf{Z0}, we obtain two
realizations of \eqrf{ZN}:
        \bea
\vc\invp\tpe\int_M\lv\ep{\mu}                      \nno
\eq\frn\sump\frac{(-\smp)^n}{\pap}\inti\de{s}s^{n-1}\ep{(s+|\mu(p)|)}   \nno
\vc+\sumk\eps^k\int_{M_0}\lvl{0}{n-1-2k}\wedge\frac{F_0^{2k}}{2^kk!}
        \eqlb{2+}     \\
\eq\frn\sump\frac{(\smp)^n}{\pap}\inti\de{s}s^{n-1}\ep{(s-|\mu(p)|)}    \nno
\vc-\sumk\eps^k\int_{M_0}\lvl{0}{n-1-2k}\wedge\frac{F_0^{2k}}{2^kk!}.
        \eqlb{2-}
        \eea

\eqrf{2+} implies that the integral \eqrf{Z} is a
polynomial in $\eps$ of degree no more than $\hfn=[\frac{\dim M_0}{4}]$
plus terms which decay not slower than $\ep{\del}$
($\del=\min\{|\mu(p)|, p\in F\}$) as $\eps\to+0$.
This coinsides with the result of Witten [\ref{W}].
If $\mi{0}$ is empty, then the integral over $M_0$ is understood to be
zero, hence the polynomial part is absent.
The novelty here is that there are two distinct formulas \eqrf{2+}
and \eqrf{2-}.
Comparing their asymptotic behavior as $\eps\to+0$, we get
	\be
\int_{M_0}\om_0^{n-1-2k}\wedge F_0^{2k}=\frac{(2\pi)^{n-1}}{2}
\sump\frac{\smp}{\pap}\mu(p)^{n-1-2k}.			\eqlb{2coh}
        \ee
In particular, take $k=0$, then
	\be
\vol(M_0)=\hf\frn\sump\frac{\smp}{\pap}\mu(p)^{n-1}.
	\ee
The significance of these formulas will be discussed in \sect{5}.

	\SECT{3}{The critical set of moment maps}

Recall that if $(M, \om)$ is a $2n$ dimensional symplectic manifold
on which there is a Hamiltonian $S^1$ action with the \mm
$\mu\colon M\to\RE$, then the critical set of $I=\mu^2$ is the union
of the critical set of $\mu$ and the zero set of $\mu$.
In this section and the next, we study the role of
these two parts respectively.

Assume temporarily that the $S^1$ action has a discrete fixed point set $F$
and the weights of the isotropic representation of $S^1$ on the tangent space
$T_pM$ ($p\in F$) are $a_i(p), i=1, \cdots, n$.
The exact stationary phase formula of \dh [\ref{DH}] is
	\be
\int_M\lv\e^{\ipm}=\frac{(2\pi)^n}{\ipn}\sump\frac{\e^{\ipm(p)}}{\pap},
	\eqlb{dh}
	\ee
as a formal series in $\phi$ or as a function of $\phi\in\RE$ or $\CO$.\\

\thm{Lemma 3.1}{Under the above assumptions,
	\bea
   \invp\tpe\int_M\lv\ep{\mu}
\eq\frn\sump\frac{(-1)^n}{\pap}\inti\de{s}s^{n-1}\ep{(s+\mu(p))}\eqlb{31+} \\
\eq\frn\sump\inv{\pap}\inti\de{s}s^{n-1}\ep{(s-\mu(p))}.	\eqlb{31-}
	\eea}

\pf{Using the \dh formula \eqrf{dh},
	\bea
\vc\invp\tpe\int_M\lv\ep{\mu}			\nno
\eq\invp\int_M\lv\int_\re\df\e^{-\hfe\phi^2+\ipm}	\nno
\eq\invp\int_\re\df\e^{-\hfe\phi^2}\int_M\lv\e^{\ipm}
   \hspace{10em}\expl{Fubini's theorem}			\nno
\eq(2\pi)^{n-1}\int_\re\df\sump\frac{\e^{-\hfe\phi^2+\ipm(p)}}{\ipn\pap}.
	\eqlb{Fou}
	\eea
Each summand has a pole at $\phi=0$. Therefore its integration over
$\phi\in\RE$ is not well defined\footnote{Alternatively, \eqrf{Fou}
is the Fourier transform of $\frac{\e^{-\hfe\phi^2}}{\ipn}$,
which is the $n$-fold integration of that of $\e^{-\hfe\phi^2}$.
There is however an ambiguity in the integration constants.}.
Nevertheless, we can deform the contour in the complex plane from
$\RE=\{\im\phi=0\}$ to $C_\pm=\{\im\phi=\pm\kap\}$ by a small number $\kap>0$
without changing the equality of \eqrf{Fou},
and then take the limit $\kap\to+0$.
The contribution of each point $p\in F$ in \eqrf{ZN} depends on the choice
$C_\pm$, but their sum does not, as long as the same choice is made for each
$p\in F$. For $\phi\in C_\pm$, we have
	\be
\inv{(i\phi)^n}=\frac{(\mp 1)^n}{(n-1)!}\inti\dr s\,s^{n-1}\,\e^{\pm i\phi s}.
	\eqlb{Schw}
	\ee
(The sign is chosen such that the integral is convergent.)
Hence in the $\kap\to+0$ limit,
	\bea
\vc(2\pi)^{n-1}\int_{C_\pm}\df\frac{\e^{-\hfe\phi^2+\ipm(p)}}{\ipn\pap}	\nno
\eq(2\pi)^{n-1}\int_{C_\pm}\df\frac{\e^{-\hfe\phi^2+\ipm(p)}}{\pap}\,
   \frac{(\mp 1)^n}{(n-1)!}\inti\dr s\,s^{n-1}\,\e^{\pm i\phi s}	\nno
\eq\frn\frac{(\mp 1)^n}{\pap}\inti\de{s}s^{n-1}\ep{(s\pm\mu(p))}.  \eqlb{31}
	\eea
The lemma follows readily.}

\thm{Proposition 3.2}{Let $(M^{2n}, \om)$ be a compact symplectic manifold
with a Hamiltonian $S^1$ action and $\mu\colon M\to\RE$, the moment map.
Assume that $S^1$ has an isolated fixed-point set $F$ and let
$a_1(p), \cdots, a_n(p)$ be the weights of isotropic representation
of $S^1$ on $T_pM$, $p\in F$. Then
        \bea
\vc\invp\tpe\int_M\lv\ep{\mu}                      \nno
\eq\frn\sump\frac{(-\smp)^n}{\pap}\inti\de{s}s^{n-1}\ep{(s+|\mu(p)|)}
+f(\eps)						\eqlb{32+}\\
\eq\frn\sump\frac{(\smp)^n}{\pap}\inti\de{s}s^{n-1}\ep{(s-|\mu(p)|)}
-f(\eps),						\eqlb{32-}
        \eea
where $f(\eps)$ is the following polynomial of degree not more than $\hfn$:
	\bea
f(\eps)\eq(2\pi)^{n-1}\sumk\frac{\eps^k}{(n-1-2k)!\,2^kk!}
          \sum_{p\in F,\;\mu(p)>0}\frac{\mu(p)^{n-1-2k}}{\pap}	\eqlb{f+}\\
       \eq-(2\pi)^{n-1}\sumk\frac{\eps^k}{(n-1-2k)!\,2^kk!}
          \sum_{p\in F,\;\mu(p)<0}\frac{\mu(p)^{n-1-2k}}{\pap}	\eqlb{f-}\\
       \eq\frac{(2\pi)^{n-1}}{2}\sumk\frac{\eps^k}{(n-1-2k)!\,2^kk!}
          \sump\frac{\smp}{\pap}\mu(p)^{n-1-2k}.		\eqlb{f}
	\eea}

\pf{Using the residue theorem, for any real number $\xi\in\RE$,
	\bea
\vc\int_{C_-}\df\frac{\e^{-\hfe\phi^2+i\phi\xi}}{(i\phi)^n}-
   \int_{C_+}\df\frac{\e^{-\hfe\phi^2+i\phi\xi}}{(i\phi)^n}		\nno
\eq i{\rm Res}_{\phi=0}\frac{\e^{\hfe(i\phi)^2+(i\phi)\xi}}{(i\phi)^n}	\nno
\eq\sumk\frac{\eps^k}{(n-1-2k)!2^kk!}\xi^{n-1-2k}.
	\eea
This, together with \eqrf{31}, implies
	\bea
\vc\inv{(n-1)!}\inti\de{s}s^{n-1}\ep{(s-\xi)}-
\frac{(-1)^n}{(n-1)!}\inti\de{s}s^{n-1}\ep{(s+\xi)}		\nno
\eq\sumk\frac{\eps^k}{(n-1-2k)!2^kk!}\xi^{n-1-2k}.		\eqlb{32}
	\eea
Formulas \eqrf{32+} and \eqrf{32-} with $f(\eps)$ given by \eqrf{f+} and
\eqrf{f-} are the results of four possible substitutions of \eqrf{32}
into \eqrf{31+} and \eqrf{31-}.
Finally, \eqrf{f} is the average of \eqrf{f+} and \eqrf{f-}.}

\eqrf{32+}, \eqrf{32-} and \eqrf{f} are in perfect agreement with
\eqrf{2+} and \eqrf{2-}.
However, we have yet to show that the coefficients in \eqrf{f} are indeed
cohomological numbers of $M_0$; this will be the task of \sect{4}.
That \eqrf{f+} and \eqrf{f-} are equal is a consequence of the proposition;
it can also be shown directly from \eqrf{dh}:

\thm{Corollary 3.3}{Under the assumptions of this section, we have
	\be
\sump\frac{\mu(p)^k}{\pap}=
\two{0}{k=0, 1, \cdots, n-1}{\frac{n!}{(2\pi)^n}\vol(M)}{k=n.}	\eqlb{Tay}
	\ee}

\pf{We expand both sides of \eqrf{dh} as formal series in $(i\phi)$ and
compare the coefficients of $(i\phi)^{k-n}$ for $k=0, 1, \cdots, n$.}

	\SECT{4}{The zero set of moment maps}

We assume that $0$ is a regular value of $\mu$.
The set $\mi{0}$ is then a connected compact submanifold of $M$ [\ref{A}]
on which the $S^1$ action is locally free.
There is a number $\del>0$ such that $\di$ consists of regular values only.
We further assume that we can choose $\del$ such that the (finite) stabilizer
subgroup $\Gam_\xi$ of $S^1$ on $\mi{\xi}$ is constant for $\xi\in\di$.
The \mm
	\be
\mu\colon\mi{\di}\to\di		\eqlb{fib}
	\ee
is then a fibration, with fibers diffeomorphic to $\mi{0}$.
There is also an $S^1$-equivariant projection
	\be
\eta\colon\mi{\di}\to\mi{0}
	\ee
defined as follows [\ref{DH}]: Choose an $S^1$-invariant connection of the
bundle \eqrf{fib},
then points on each horizontal curve have the same projection under $\eta$.
We thus have a diffeomorphism
	\be
(\eta,\mu)\colon\mi{\di}\to\mi{0}\times\di	\eqlb{diff}
	\ee
and the induced $S^1$ action on the second component $\di$ is trivial
[\ref{DH}].
For each $\xi\in\di$, the restriction of $\eta$ to $\mi{\xi}$ is an
$S^1$-equivariant diffeomorphism $\eta_\xi\colon\mi{\xi}\to\mi{0}$.
Hence the symplectic quotient $M_\xi=\mi{\xi}/S^1$ can be identified
with $M_0=\mi{0}/S^1$.

On the other hand, each level set $\mi{\xi}$, $\xi\in\di$, is the total
space of a principal $S^1/\Gam_\xi$-bundle over the quotient $M_\xi$.
Let $X$ be the vector field on $M$ induced by the $S^1$ action,
and $Y$, the horizontal lift of the tangent vector $\pdo{\xi}$ on $\di$.
The 1-form $\alpha=-i_Y\om$ defines a connection of the bundle
$\pi\colon\mi{\xi}\to M_\xi$ for each $\xi$ [\ref{DH}], since $\alpha$
is $S^1$-invariant and $i_X\alpha=-i_Xi_Y\om=i_Y\dr\mu=L_Y\mu=1$.
The integration of $\alpha$ along each $S^1/\Gam_\xi$-fiber is
$\frac{2\pi}{|\Gam_\xi|}=\frac{2\pi}{|\Gam_0|}$.
Both the canonical symplectic form $\om_\xi$ and the curvature of
$\pi\colon\mi{\xi}\to M_\xi$ are 2-forms on $M_\xi$ and can be viewed as
2-forms on $M_0$ through the identification of $M_\xi$ with $M_0$.
It can be shown that [\ref{DH}]
	\be
\pdo{\xi}\om_\xi=F_\xi.		\eqlb{var}
	\ee
Since $F_\xi$ is the curvature of a continuous family of bundles,
the cohomology class $[F_\xi]\in H^2(M_0)$ is a constant.
Hence \eqrf{var} implies [\ref{DH}]
	\be
[\om_\xi]=[\om_0]+\xi\,[F_0].	\eqlb{sol}
	\ee

\thm{Lemma 4.1}{Under the diffeomorphism \eqrf{diff},
the symplectic form on $\mi{\di}$ is
	\be
\om=\alpha\wedge\dr\xi+\pi^*\om_\xi.		\eqlb{4om}
	\ee}

\pf{One need to check that the two sides are equal on $X$, $Y$ and
the horizontal vector fields of the bundle $\mi{\xi}\to M_\xi$.
This follows from $i_X\om=\dr\mu$, $i_Y\om=-\alpha$ and
the definition of $\om_\xi$.}

\thm{Proposition 4.2}{Let $M$ be a compact symplectic manifold
of dimension $2n$, on which there is a Hamiltonian $S^1$ action
with the \mm $\mu\colon M\to\RE$.
Assume that $0$ is a regular value of $\mu$ and the stabilizer is
a constant (finite) subroup $\Gam_0\subset S^1$.
Let $M_0=\mi{0}/S^1$ be the symplectic quotient with
the canonically induced symplectic structure $\om_0$.
Choose a connection of the principal $S^1/\Gam_0$-bundle
$\mi{0}\to M_0$ with the curvature 2-form $F_0\in\Om^2(M_0)$.
Then there is a number $\del>0$ such that
	\be
\invp\tpe\int_M\lv\ep{\mu}
=\invg\sumk\eps^k\int_{M_0}\lvl{0}{n-1-2k}\wedge\frac{F_0^{2k}}{2^kk!}+\err.
	\eqlb{4+}
	\ee}

\pf{Since $M$ is compact,
	\be
\invp\tpe\int_M\lv\ep{\mu}=\invp\tpe\int_{\mi{\di}}\lv\ep{\mu}+\err.
	\ee
On the other hand,
	\newpage
	\bea
\vc\invp\tpe\int_{\mi{\di}}\lv\ep{\mu}				\nno
\eq\invp\tpe\int_{\mi{0}\times\di}
   \frac{(\alpha\wedge\dr\xi+\pi^*\om_\xi)^n}{n!}\ep{\xi}
   \hspace{8.6em}\expl{using \eqrf{4om}}			\nno
\eq\inv{|\Gam_\xi|}\int_{-\del}^\del\de{\xi}\ep{\xi}\int_{M_\xi}\lvl{\xi}{n-1}
   \hspace{7em}\expl{integration along the fibers}		\nno
\eq\invg\int_{-\del}^\del\de{\xi}\ep{\xi}
   \int_{M_0}\frac{(\om_0+\xi F_0)^{n-1}}{(n-1)!}
   \hspace{11.3em}\expl{using \eqrf{sol}}				\nno
\eq\invg\sumk\int_{-\del}^\del\de{\xi}\xi^{2k}\ep{\xi}
   \int_{M_0}\lvl{0}{n-1-2k}\wedge\frac{F_0^{2k}}{(2k)!}	\nno
\eq\invg\sumk\intii\de{\xi}\xi^{2k}\ep{\xi}
   \int_{M_0}\lvl{0}{n-1-2k}\wedge\frac{F_0^{2k}}{(2k)!}+\err	\nno
\eq\invg\sumk(2k-1)!!\eps^k
   \int_{M_0}\lvl{0}{n-1-2k}\wedge\frac{F_0^{2k}}{(2k)!}+\err	\nno
\eq\invg\sumk\eps^k\int_{M_0}\lvl{0}{n-1-2k}\wedge\frac{F_0^{2k}}{2^kk!}+\err.
	\eea }

So, the polynomial coefficients are indeed the cohomological pairings
$\langle[\om_0]^{n-1-2k}\cup[F_0]^{2k}, [M_0]\rangle$ on $M_0$.
Here we made no assumptions on the critical set of $\mu$.
It should also be noted that if $G=T$ is a torus group, the above argument
remains valid if we replace $(\th, \xi)$ by a multi-component coordinate.
The polynomial coefficients are pairings of the quotient symplectic form
with various components of the curvature valued in the Lie algebra of $T$.

        \SECT{5}{Theorems and Applications}

We know combine the results of the previous two sections.

\thm{Theorem 5.1}
{Let $(M^{2n}, \om)$ be a compact symplectic manifold with a Hamiltonian
$S^1$ action and $\mu\colon M\to\RE$, the moment map.
Assume that $S^1$ has an isolated fixed-point set $F$ and let
$a_1(p), \cdots, a_n(p)$ be the weights of isotropic representation
of $S^1$ on $T_pM$, $p\in F$.
Assume also that $0$ is a regular value of $\mu$ and the stabilizer of
the $S^1$ action on a neighborhood of $\mi{0}$ is of constant type $\Gam_0$.
Let $M_0=\mi{0}/S^1$ be the symplectic quotient with the canonically
induced symplectic structure $\om_0$.
Choose a connection of the principal $S^1$-bundle $\mi{0}\to M_0$
with the curvature 2-form $F_0\in\Om^2(M_0)$. Then
        \bea
\vc\invp\tpe\int_M\lv\ep{\mu}			                        \nno
\eq\frn\sump\frac{(-\smp)^n}{\pap}\inti\de{s}s^{n-1}\ep{(s+|\mu(p)|)}   \nno
\vc+\invg\sumk\eps^k\int_{M_0}\lvl{0}{n-1-2k}\wedge\frac{F_0^{2k}}{2^kk!}
	\eqlb{51+}	\\
\eq\frn\sump\frac{(\smp)^n}{\pap}\inti\de{s}s^{n-1}\ep{(s-|\mu(p)|)}    \nno
\vc-\invg\sumk\eps^k\int_{M_0}\lvl{0}{n-1-2k}\wedge\frac{F_0^{2k}}{2^kk!}.
	\eqlb{51-}
        \eea}

\pf{By comparing the terms in \eqrf{32+} and \eqrf{4+} that are
polynomial in $\eps$, we get
        \be
\int_{M_0}\om_0^{n-1-2k}\wedge F_0^{2k}=\frac{|\Gam_0|}{2}(2\pi)^{n-1}
\sump\frac{\smp}{\pap}\mu(p)^{n-1-2k},			\eqlb{5coh}
        \ee
and that $\del$ in Proposition 4.2 can be as big as $\{|\mu(p)|, p\in F\}$.
\eqrf{51+} and \eqrf{51-} then follow from \eqrf{32+}, \eqrf{32-}
and \eqrf{5coh}.}

This theorem confirms formulas \eqrf{2+} and \eqrf{2-} derived in \sect{2}.
Moreover, \eqrf{5coh} is an interesting formula about
the cohomological numbers of the symplectic quotient $M_0$.
Since $\mu-\xi$ is just another \mm for any $\xi\in\RE$, we have,
taking $k=0$ in \eqrf{5coh}, the following

\thm{\bf Corollary 5.2}
{Under the assumptions of Theorem 5.1, except that $0$ is replaced
by any regular value $\xi\in\RE$,
the Liouville volume of the symplectic quotient $M_\xi=\mi{\xi}/S^1$ is
        \be
\vol(M_\xi)=\frac{|\Gam_\xi|}{2}\frn
\sump\frac{\sgn(\mu(p)-\xi)}{\pap}(\mu(p)-\xi)^{n-1}.	\eqlb{5vol}
        \ee}
\noindent $\Box$ \\

\noindent
It was known that $\vol(M_\xi)$ is a piecewise polynomial in $\xi$ [\ref{DH}].
\eqrf{5vol} provides  an explicit formula,
which seems not to have been obtained by other methods.
Part of Corollary 3.3 (when $k<n$) follows from
${\displaystyle \lim_{\xi\to\infty}\vol(M_\xi)=0}$
by collecting the coefficients of $\xi^k$ in \eqrf{5vol}.
Positivity of $\vol(M_\xi)$ when $\xi$ is finite may impose additional
constraints on $a_i(p)$ and $\mu(p)$.

In the other extreme, if $\dim M_\xi$ is a multiple of $4$, i.e., $2(n-1)=4m$
for $m\in\NA$, then taking $k=m$ in \eqrf{5coh}, we have
        \be
\int_{M_\xi}\lo\frac{F_\xi}{2\pi}\lc\ind{2m}=\frac{|\Gam_\xi|}{2}
\sump\frac{\sgn(\mu(p)-\xi)}{\prod_{i=1}^{2m+1}a_i(p)}.		\eqlb{top}
        \ee
Let $c_1(F_\xi)$ be the first Chern class of the circle bundle, then
the left hand side is the pairing $\langle c_1(F_\xi)^{2m}, [M_\xi]\rangle$.
Topological invariance is confirmed since the right hand side
is a piecewise constant function, which jumps only at points
where topological type of $M_\xi$ or that of the circle bundle changes.
As a corollary, the right hand side is an integer for any regular value $\xi$.
(That it is rational is obvious.)
It remains open whether \eqrf{51+} and \eqrf{51-} have other applications
to symplectic geometry.

So far, except in \sect{4}, we assumed that the fixed-point set $F$
is discrete.
If there are connected components of $F$ of dimensions higher than zero,
then each such component $N$ is a symplectic submanifold with the pull-back
symplectic structure $\om_N$.
The \mm has a constant value on $N$, denoted by $\mu(N)$.
The \dh formula should be modified to [\ref{DH}]
	\be
\int_M\lv\e^{\ipm}=\frac{(2\pi)^n}{\ipn}\sump\frac{\e^{\ipm(p)}}{\pap}
+\sumn\int_N\frac{\e^{i\phi\mu(N)}\,\e^{\om_N}}{\euler}.
	\ee
Here the second sum is over the connected components $N\subset F$ of $\dim N>0$
and $\euler$ is the equivariant Euler class of the normal bundle $\nu_N$
of $N$.
Let $n_N=n-\hf\dim N$ and $a_i(N)$ ($i=1, \cdots, n_N$) be the weights
of the $S^1$ action on the fibers of $\nu_N$.	Then [\ref{DH}]
	\be
\inv{\euler}=\frac{(2\pi)^{n_N}}{\pan}\sum_{k=0}^{\hf\dim N}
					\inv{(i\phi)^{n_N+k}}\beta_k(N),
	\ee
where $\beta_k(N)$ are $2k$-forms on $N$ explicable in terms of the
$S^1$-invariant curvature $F_N$ of $\nu_N$, and $\beta_0(N)=1$.
Therefore
	\be
\int_N\frac{\e^{i\phi\mu(N)}\,\e^{\om_N}}{\euler}
=\frac{(2\pi)^{n_N}}{\pan}\sumkn\frac{\e^{i\phi\mu(N)}}{(i\phi)^k}
\int_N\lvl{N}{n-k}\wedge\beta_{k-n_N}(N).
	\ee
We are thus in a position analogous to \eqrf{Fou}.
Using \eqrf{Schw} and after computations similar to those in \sect{3}, we have

\thm{Theorem 5.3}{If the fixed-point set $F$ contains connected components
$N$ of $\dim N>0$, then under the above notations, formula \eqrf{51+}
(and \eqrf{51-} respectively) shall be modified by adding
	\be
\sumn\frac{(2\pi)^{n_N-1}}{\pan}\inti\de{s}\lo\sumkn
\frac{(\mp\sgn\mu(N))^k}{(k-1)!}\int_N\lvl{N}{n-k}\wedge\beta_{k-n_N}(N)
\,s^{k-1}\lc\,\ep{(s\pm|\mu(N)|)}.
	\ee}
\noindent $\Box$ \\

\noindent
The terms that modify \eqrf{51+} are exponentially small as $\eps\to+0$.
We can also generalize \eqrf{5vol} by adding an extra piecewise polynomial
	\be
\frac{|\Gam_\xi|}{2(n-1)!}\sumn\frac{(2\pi)^{n_N-1}}{\pan}\sumkn\int_N
\lvl{N}{n-k}\wedge\beta_{k-n_N}(N)\sgn(\mu(N)-\xi)(\mu(N)-\xi)^{k-1}. \eqlb{vn}
	\ee
It is in general discontinuous at $\xi=\mu(N)$ if $N$ is of codimension 2
in $M$.

	\SECT{6}{Examples}

We begin with the example considered by Witten [\ref{W}].
Let $M=S^2$ with the standard symplectic form $\om=\dr\psi\wedge\dr\cos\th$ in
the spherical coordinates $(\th, \psi)$, ($0\le\th\le\pi$, $0\le\psi\le 2\pi$).
The group $S^1$ acts on $M$ as rotations with the fixed points
$N$ ($\th=0$) and $S$ ($\th=\pi$).
This action is Hamiltonian and the \mm is $\mu=\cos\th$.
For $\xi\in\RE$, the level set $\mi{\xi}$ is empty if $|\xi|>1$ and is a circle
$\th=\cos^{-1}\xi$ if $|\xi|<1$.
The symplectic quotient $M_\xi=\mi{\xi}/S^1$ is therefore
	\be
M_\xi=\two{{\mbox{a point}}}{|\xi|<1}{\emptyset}{|\xi|>1.}   \eqlb{pt}
	\ee
The formula \eqrf{51+} for the \mm $\mu-\xi$ reduces to
(after the integration over $\psi$ and a change of variable
$\cos\th=s$ on the left hand side) [\ref{W}]
	\bea
\vc\int_{-1}^1\de{s}\ep{(s-\xi)}	\nno
\eq\three{1-\int_0^\infty\de{s}\ep{(s+1-\xi)}
	  -\int_0^\infty\de{s}\ep{(s+1+\xi)}}{-1<\xi<1}
	 {\int_0^\infty\de{s}\ep{(s-1+\xi)}-\int_0^\infty\de{s}\ep{(s+1+\xi)}}
	 {\xi>1}
	 {\int_0^\infty\de{s}\ep{(s-1-\xi)}-\int_0^\infty\de{s}\ep{(s+1-\xi)}}
	 {\xi<-1.}
	\eea
Formula \eqrf{5vol} reduces to
	\be
\vol(M_\xi)=\two{1}{|\xi|<1}{0}{|\xi|>1;}
	\ee
this is clearly consistent with \eqrf{pt}.

Next, consider the product $M=S^2\times S^2$ with the diagonally induced
$S^1$ action.
Under two spherical coordinates $(\th_1, \psi_1, \th_2, \psi_2)$,
the symplectic form is $\om=\dr\psi_1\wedge\dr\cos\th_1
+\dr\psi_2\wedge\dr\cos\th_2$ and the \mm is $\mu=\cos\th_1+\cos\th_2$.
The critical points of $\mu$ is $(N,N)$, $(N,S)$, $(S,N)$, $(S,S)$,
on which $\mu$ takes values $2$, $0$, $0$, $-2$ respectively.
The level set $\mi{\xi}$ is empty if $|\xi|>2$;
if $0<|\xi|<2$, it is diffeomorphic to $S^3$ for the following reason.
For $0<\xi<2$, $\mi{\xi}$ can be mapped onto
$S^3=\{(z_1,z_2)\in\CO^2, |z_1|^2+|z_2|^2=1\}$ by
	\be
z_k=\lo\frac{1-\cos\th_k}{2-\xi}\lc\ind{\hf}
\,\e^{i\psi_k}, \;\;\; (k=1,2).\eqlb{onto}
	\ee
$S^1$ acts on $S^3$ freely by changing a common phase of $z_1$ and $z_2$.
The symplectic quotient is the standard Hopf fibration
$\mi{\xi}=S^3\to M_\xi=S^2$. The case $-2<\xi<0$ is identical.
Formula \eqrf{5vol} yields
	\be
\vol(M_\xi)=\two{0}{|\xi|>2}{2\pi(2-|\xi|)}{|\xi|<2.}
	\ee
Though the Hopf fibration is non-trivial, the volume of $M_\xi$, if
non-vanishing, is the only term in \eqrf{51+} that does not fall as $\err$
when $\eps\to+0$.
This is a feature of all symplectic 4-manifolds admitting Hamiltonian
$S^1$ actions.

Consider now the $n$-fold product $M=S^2\times\cdots\times S^2
(n {\mbox{ copies}})=\{\,(\th_k,\psi_k), \; k=1,\cdots,n\}$.
Again, $S^1$ acts diagonally by the rotation on each component.
The fixed points are $(x_1,\cdots,x_n)$ $\in M$ with $x_k=N$ or $S$.
The \mm $\mu=\sum_{k=1}^n\cos\th_k$ has critical value $n-2p$
at $(x_1,\cdots,x_n)$ if there are $p$ $x_k$'s equal to $S$.
Formulas \eqrf{5vol} and \eqrf{top} now reduce to
	\be
\vol(M_\xi)=\hf\frn\sum_{p=0}^n(-1)^p\comb{n}{p}\sgn(n-2p-\xi)(n-2p-\xi)^{n-1}
	\ee
and (if $n-1=2m, m\in\NA$)
	\be
\int_{M_\xi}\lo\frac{F_\xi}{2\pi}\lc\ind{2m}
=\hf\sum_{p=0}^n(-1)^p\comb{n}{p}\sgn(n-2p-\xi).
	\ee
If $n-2<|\xi|<n$, the same trick of \eqrf{onto} shows that
$\mi{\xi}=S^{2n-1}$, and its quotient onto $M_\xi=\CO P^{n-1}$
is a (generalized) Hopf fibration.
We have $\vol(M_\xi)=\frn(n-|\xi|)^{n-1}$ and if $\dim M_\xi=4m$,
$\int_{M_\xi}(\frac{F_\xi}{2\pi})^{2m}=1$.
Alternatively, cohomological dual of $[M_\xi]$ is the
top wedge product of the first Chern class $[\frac{F_\xi}{2\pi}]$
of the tautological bundle, hence the last equality.
Since $\vol(M)=(4\pi)^n$, the equalities \eqrf{Tay} are
	\be
\sum_{p=0}^n(-1)^p\comb{n}{p}(n-2p-\xi)^k=\two{0}{0\le k\le n-1}{2^nn!}{k=n.}
	\ee
They are consequences of the well-known identities
	\be
\sum_{p=0}^n(-1)^p\comb{n}{p} p^k=\two{0}{0\le k\le n-1}{(-1)^nn!}{k=n.}
	\ee

Finally, we study the integration \eqrf{Z} for symplectic 4-manifolds
(i.e., $n=2$) with a Hamiltonian action.
Proposition 4.2 implies that the polynomial part is $\pm\invg\vol(M_0)$,
where $M_0=\mi{0}/S^1$ is the (two dimensional) quotient.
The rest consists of contributions from critical points $p\in F$
and critical surfaces $N\subset F$.
The former can be obtained by Theorem 5.1.
Each critical surface $N$ has $\dim N=2$, hence $n_N=1$.
The Morse index (in the sense of Bott) of $N$, being even, may only be 0 or 2.
So the \mm is either locally maximal or minimal on $N$, in which case
the weight of the $S^1$ action $a_1(N)$ is positive or negative respectively.
The complexification of its normal bundle has the form
$\nu^\co_N=l_N\oplus l^*_N$, where $l_N$ is a complex line bundle over $N$.
Let $F_N$ be the curvature of an $S^1$-invariant connection on $l_N$,
then $\inv{\euler}=\frac{2\pi}{i\phi}(1-\inv{i\phi}F_N)$,
so $\beta_0(N)=1$, $\beta_1(N)=-F_N$. To summarize,
	\bea
\!\!\!\vc\inv{2\pi}\tpe\int_M\frac{\om^2}{2}\,\ep{\mu}=
   2\pi\sump\app\inti\de{s}s\,\ep{(s\pm|\mu(p)|)}			\nno
\sep{+}\sumn\ann\inti\de{s}(\mp\sgn\mu(N)\vol(N)-2\pi c_1(\nu_N)\,s)
   \,\ep{(s\pm|\mu(N)|)}\pm\invg\vol(M_0),				\nno
	\eea
where $c_1(\nu_N)=\langle[\frac{F_N}{2\pi}], [N]\rangle\in\ZE$ is the first
Chern number of $l_N$.
\eqrf{5vol}, modified by \eqrf{vn}, yields
	\bea
\!\!\!\vc\vol(M_\xi)=\pi|\Gam_\xi|\sump\app|\mu(p)-\xi|		\nno
\sep{+}|\Gam_\xi|\sumn\ann\lo\hf\sgn(\mu(N)-\xi)\vol(N)
	-\pi c_1(\nu_N)|\mu(N)-\xi|\lc.				\eqlb{6vol}
	\eea
This is a relation about, among other things, the volume (or more precisely,
the area) of the quotient and those of the critical surfaces.
Since $M_\xi=\emptyset$ as $\xi\to\infty$, collecting the coefficients of
the linear and constant terms in \eqrf{6vol}, we get
	\be
\sump\app=\sumn\frac{c_1(\nu_N)}{a_1(N)}	\eqlb{6con}
	\ee
and
	\be
\sump\app\mu(p)+\sumn\ann\lo\frac{\vol(N)}{2\pi}-c_1(N)\mu(N)\lc=0 \eqlb{6lin}
	\ee
respectively. These are the actually formula \eqrf{Tay} ($k=0, 1$) modified
by the presence of critical surfaces.

It was known that if the action of $S^1$ is semi-free\footnote{An action
is semi-free if it is free outside the fixed set.} and if all the
fixed points are isolated, then there are at least four [\ref{Au}].
This result also follows immediately from \eqrf{6con}, since
$a_i(p)=\pm 1$ and $a_1(p)a_2(p)=1$ if $\mu$ is extremal at $p$.
A typical example is the 4-manifold $S^2\times S^2$ considered above.
If there are only two critical values, the maximum and the minimum can be
reached both on a surface of the same topological type, or one is reached
on an isolated fixed point, the other, on a sphere [\ref{Au}].
An example of the latter case is $\CO P^2=\{[x, y, z]\}$ with the standard
K\"ahler form.
The action of $S^1=U(1)$ is $[x,y,z]\mapsto[x,y,tz]=[t^{-1}x,t^{-1}y,z]$
for $t\in U(1)$.
The \mm is $\mu([x,y,z])=\hf\frac{|z|^2}{|x|^2+|y|^2+|z|^2}$ [\ref{Au}].
It reaches maximum at the point $p=[0,0,1]$ with $\mu(p)=\hf$ and
$a_1(p)a_2(p)=-1$, and minimum at the sphere $N=\{[x,y,0]\}$ with
$\mu(N)=0$ and $a_1(N)=-1$.
\eqrf{6con} and \eqrf{6lin} can be used to solve $c_1(\nu_N)=-1$
and $\vol(N)=\pi$. Consequently,
	\be
\vol(M_\xi)=\two{\pi(1-2\xi)}{0<\xi<\hf}{0}{\xi<0\mbox{ or }\xi>\hf.}
	\ee
It would be interesting to explore possible relations between these formulas
and the classification of symplectic 4-manifolds with a Hamiltonian
$S^1$ action [\ref{Au}].

        \newcommand{\rf}[1]{\item \label{#1}}

        \newcommand{\athr}[2]{{#1}.$\,${#2}}
        \newcommand{\au}[2]{\athr{{#1}}{{#2}},}
        \newcommand{\an}[2]{\athr{{#1}}{{#2}} and}

        \newcommand{\jr}[6]{{\it {#1}}, {#2} {#3} ({#4}) {#5}-{#6}.}

        \newcommand{\pr}[3]{{\it {#1}}, {#2} ({#3}).}

        \newcommand{\bk}[4]{{\it {#1}} ({#2}, {#3}, {#4})}

        \newcommand{\cf}[8]{{\it {#1}}, in: {\it {#2}}, {#5}, pp.$\,${#3}-{#4}
                 ({#6}, {#7}, {#8}).}

        \vspace{5ex}
        \begin{center}
{\bf References}
        \end{center}
{\small
        \newcounter{rfs}
        \begin{list}%
        {[\arabic{rfs}]}{\usecounter{rfs}}

	\rf{A}
	\au{M.$\,$F}{Atiyah}
	\jr{Convexity and commuting Hamiltonians}
	{Bull. London Math. Soc.}{14}{1982}{1}{15}

        \rf{AB}
        \an{M.$\,$F}{Atiyah} \au{R}{Bott}
        \jr{The moment map and equivariant cohomology}
        {Topology}{23}{1984}{1}{28}

	\rf{Au}
	\au{M}{Audin}
	\cf{Hamiltoniens p\'eriodiques sur les vari\'et\'e symplectiques
	    compactes de dimension 4}
	   {G\'eom\'etrie symplectique et m\'echanique, Proceedings 1988}
	   {1}{25}{C.$\,$Albert ed., Lecture Notes in Mathematics 1416}
	   {Springer}{Berlin, Heidelberg, New York}{1990}

        \rf{BV}
        \an{N}{Berline} \au{M}{Vergne}
        \jr{Z\'ero d'un champ de vecteurs et classes caract\'eristiques
        \'equivariantes}{Duke Math. J.}{50}{1983}{539}{549}

        \rf{DH}
        \an{J.$\,$J}{Duistermaat} \au{G.$\,$J}{Heckman}
        \jr{On the variation in the cohomology of the symplectic form of the
        reduced phase space}{Invent. Math.}{69}{1982}{259}{268;
        Addendum, {\it ibid.} 72 (1983) 153-158}

        \rf{GP}
        \an{V}{Guillemin} \au{E}{Prato}
        \jr{Heckman, Kostant, and Steinberg formulas for symplectic manifolds}
        {Adv. Math.}{82}{1990}{160}{179}

        \rf{MQ}
        \an{V}{Mathai} \au{D}{Quillen}
        \jr{Superconnections, Thom classes, and equivariant differential forms}
        {Topology}{25}{1986}{85}{110}

        \rf{W}
        \au{E}{Witten}
        \jr{Two dimensional gauge theory revisited}
	{J. Geom. Phys.}{9}{1992}{303}{368}
        \end{list}      }

        \end{document}